\documentclass[traditabstract]{aa}
\usepackage{graphicx,natbib,color}
\usepackage{txfonts,afterpage,hyperref}

\def\lta{ \lower .75ex\hbox{$\sim$} \llap{\raise .27ex \hbox{$<$}} }

\begin{document}

\date{Received .../Accepted ...}

\title{Simultaneous mm/X-ray intraday variability in the radio-quiet AGN MCG+08-11-11}
\author{P.-O. Petrucci\inst{1}
 \and V. Pi\'etu\inst{2}
  \and E. Behar\inst{3}
  \and M. Clavel\inst{1}
  \and S. Bianchi\inst{4}
\and G. Henri\inst{1}
\and S. Barnier\inst{5}
  \and S. Chen\inst{3}
   \and J. Ferreira\inst{1}
 %  \and et al.\inst{}
%  \and G. Matt\inst{4}
%  \and M. Dadina\inst{2}
\and J. Malzac\inst{6}
  \and R. Belmont\inst{7}
  \and S. Corbel\inst{7,8}
  \and M. Coriat\inst{6}
  }
  
\institute{Univ. Grenoble Alpes, CNRS, IPAG, F-38000 Grenoble, France 
	   \and
	   IRAM, 300 rue de la piscine, F-38406 Saint Martin d'Heres, France
	   \and 
	   Department of Physics, Technion, Haifa 32000, Israel
	   \and
	   Dipartimento di Matematica e Fisica, Universit\`a degli Studi Roma Tre, 
	   via della Vasca Navale 84, 00146 Roma, Italy 
	   \and
	   Department of Earth and Space Science, Osaka University, 560-0043, Japan
	    \and
	   IRAP, Universit\'e de toulouse, CNRS, UPS, CNES , Toulouse, France
	\and
	    ORN, Observatoire de Paris, CNRS, PSL, Universit\'e d'Orl\'eans, Nançay, France
     \and AIM, CEA, CNRS, Universit\'e Paris Cit\'e, Universit\'e Paris-Saclay, F-91191 Gif-sur-Yvette, France
} 

%\author{P.O.\,Petrucci\inst{1} et al.}
%
%\institute{UJF-Grenoble 1 / CNRS-INSU, Institut de Plan¬¨√©tologie et d'Astrophysique de Grenoble (IPAG) UMR 5274, Grenoble, F-38041, France }

\abstract{Most of the Active Galactic Nuclei (AGN) are radio-quiet (RQ) and, differently from radio-loud (RL) AGN, do not show signature of large-scale and powerful jets. The physical origin of their radio emission remains then broadly unclear. The observation of flat/inverted radio spectra at GHz frequencies seems to support however the presence of an unresolved synchrotron self-absorbed region in the close environment of the supermassive black hole. 
%If true, its size should be of the order of the X-ray corona one when observed in the millimetric (mm) domain. 
Its size could be as small as that of the X-ray corona. Since synchrotron self absorption decreases strongly with frequency, these sources need to be  observed in the millimetric (mm) domain. We report here a 12h simultaneous mm-X-ray observation of the RQ AGN MCG+08-11-11 by NOEMA and NuSTAR, respectively.  
The mm flux shows a weak but clear increase along the pointing with %significant variability characterized by 
a fractional variability of $2.0\pm0.1$\%.  
%On the same period 
The 3-10 keV flux of NuSTAR also increases and shows a fractional variability of $7.0\pm1.5$\%. A structure function analysis shows a local maximum in the mm light curve corresponding to 2-3\% of variability on timescale of $\sim2\times10^4$ seconds  (100-300 $R_g$ light crossing time). Assuming an optically thick mm emitting medium, this translates into an upper limit of its size of $\sim$1300 $R_g$.
The observation of fast variability in radio-mm and X-ray wavelengths, as well as a similar variability trend, well support the mm emission to be emitted by a region close, and potentially related to, the X-ray corona like an outflow/weak jet. %This also put physical constraints on the emitting plasma (size, magnetic field strength, presence of non-thermal particles). 
}

%Black hole physics -- Accretion, accretion discs -- Magnetohydrodynamics (MHD) -- ISM: jets and outflows -- X-rays: binaries}

\maketitle

%%%%%%%%%%%%%%%%%%%%%%%%%%%%%%%%%%%%%%%%%%%%%%%%%%%%%%%%%%%%
\section{Introduction}
%%%%%%%%%%%%%%%%%%%%%%%%%%%%%%%%%%%%%%%%%%%%%%%%%%%%%%%%%%%%
 \label{intro}
Active Galactic Nuclei (AGN) can be subdivided in two main groups, the Radio-Quiet (RQ) and Radio-Loud (RL) AGN, depending on their radio-loudness $R_{\nu}$\footnote{$R_{\nu}$ is generally defined as the ratio between the rest frame radio luminosity at a given frequency $\nu$ and the optical luminosity, usually in the B band.} \citep{kel89}.  The threshold value between the two groups is typically $R_{5GHz}\sim10$. 
The two groups are also characterised by a different radio to X-ray luminosity ratio \citep{Terashima2003}, where $L_R/L_X\sim 10^{-2}$ for RL AGN, and $L_R/L_X\sim 10^{-5}$ in RQ AGN \citep{lao08}.
In the case of RL AGN the {origin of their radio emission} is relatively well understood and is a direct signature of synchrotron emission from large-scale and powerful jets. 
On the contrary, the physical origin of the radio emission in RQ AGN is still unclear. It is suspected to be a mix of %synchrotron 
emission of different origins. 
The most favoured ones are: nuclear star-forming regions, weak/small scale jets and/or an opaque (optically thick) unresolved source in the close environment of the accretion disk around the supermassive black hole, referred to as the corona \citep[see][for a recent review]{pane19}. 
The presence of the latter is sustained by the observation of flat/inverted radio spectra at $\sim100$\,GHz frequencies, characteristic of synchrotron self-absorption emission, 
which exceeds the low-frequency spectral slope \citep{doi11,par13,behar15,behar18,inouedoi2018}, the so called mm-wave excess. This excess luminosity can extend with a flat slope up to 230\,GHz \citep{kawamuro22}, and shows a tight $L_{mm}/L_X\sim 10^{-4}$ correlation with the X-ray luminosity \citep{behar15, behar18, kawamuro22}.

Recent observations with the VLBA, reaching a resolution of a few mas, which corresponds in some cases to a few pc, also suggest that the radio emission of low Eddington ratio RQ AGN ($L_{AGN}/L_{Edd}\lesssim$ 0.3) predominately originates from an unresolved and extremely compact region (\citealt{Alhosani22}). A larger VLBA sample shows that most RQ AGN have a flat-slope compact core that coincides with the Gaia position, and whose luminosity tightly correlates with the X-ray luminosity \citep{chen23}. Currently it is not possible to go below the pc scale with imaging, and one has to turn to variability time scales to constrain the source size based on light travel time arguments. This is the purpose of this letter, presenting for the first time intraday (hour timescale) variability of simultaneous mm (NOEMA) and X-ray (NuSTAR) observations of an AGN (MCG+08-11-11).

Simple estimates indeed show that the size of an opaque self absorbed synchrotron source decreases strongly with frequency. The physical size $R$ of a self-absorbed synchrotron source can indeed be estimated from its measured radio flux density $F_{\nu}$ (in units of $\mu Jy$) at the frequency $\nu$  (e.g. following \citealt{lao08}):

\begin{eqnarray}
R &\simeq& 2.5\times 10^{17} \left(\frac{F_{\nu}}{\mu Jy}\right)^{1/2}\nu_{GHz}^{-5/4}B_G^{1/4}z\ \mbox{cm} \label{eq1}\\
&\simeq& 1.3\times 10^{5} \left(\frac{F_{\nu}}{\mu Jy}\right)^{1/2}\nu_{GHz}^{-5/4}B_G^{1/4}zM_7^{-1}R_g\label{eq1b}
\end{eqnarray}

\noindent where $\nu_{GHz}$ is the frequency in GHz, $B_G$ is the magnetic field strength (assumed to be uniform) in Gauss and $z$ the redshift of the AGN, a proxy for its distance at low-$z$, assuming a Hubble constant equal to 70 km s$^{-1}$ Mpc$^{-1}$. Equation \ref{eq1b} is rescaled with respect to the gravitational radius $R_g$ of a $10^7 M_{\odot}$ supermassive black hole\footnote{$R_g=1.5\times 10^{12} M_7$ cm with $M_7=M/10^7 M_{\odot}$}.  % \citep{par13}. %(Park, Sohn \& Yi 2013). 
Since above 300 GHz the thermal dust emission starts to dominate \citep{bar92}, the best radio window to observe radio emission as close as possible to the central black hole is in the range $\sim$ 50-250 GHz.\\
\begin{table*}
\begin{center}
\begin{tabular}{ccccc}
\hline
Start NOEMA observation & NOEMA Exposure & Start NuSTAR observation & NuSTAR  exposure  \\
(MJD) & (h) & (MJD) & (ks on source) & \\
\hline
 %NGC 7469 & & & & & XMM\\
 %NGC 5506 & & & & & XMM\\
 59566.708333 & 13.15 & 59566.688298 & 25 &\\
 \hline
 \end{tabular}
 \end{center}
 \caption{Log of the Radio mm/X-ray observations of MCG+08-11-11 by NOEMA and NuSTAR. MJD 59566 corresponds to  Dec. the 18th 2021\label{logobs}}
 \end{table*}

Equation \ref{eq1} shows that for sources dominated by self-absorbed synchrotron emission with radio flux density at 100\,GHz in the range of 1-10\,mJy and redshift of 0.02 (which is the one of MCG+08-11-11), the size of the radio photosphere is of the order of $10^{14}-10^{16}$  cm ($\sim$1-100\,hours light crossing time). This corresponds to $10^2-10^4$ gravitational radii for a supermassive black hole of $10^7 M_{\odot}$ (see Fig. \ref{fig1} with the contour plot of the radio flux density $F_{\nu}$ in the $R-B_G$ plane). These sizes are close to the estimated size of the X-ray emitting region in RQ AGN, the so-called hot corona, a plasma of hot electrons at a temperature of kT$\sim$100\,keV estimated from hard X-ray spectra (e.g. \citealt{per02,fab15, fab17, tor18,akylas21}).
Variability of the mm emission (on $\sim$day timescale) already supports the small ($<$ light day) size of (part of) the mm radio emitting region (e.g., \citealt{doi11}). 
The similar inter-day variability parameters detected at 100\,GHz and in X-rays in the RQ AGN NGC 7469 adds even more to the evidence that the mm and X-ray emission may have the same physical origin, and could be both associated with the hot corona \citep{baldi15,behar20}. Actual inter-day temporal correlation between radio and X-ray light curves, however, is much harder to substantiate \citep{panessa22, Chen22}, mostly due to radio photometric stability over weeks and months.\\

Here, we introduce a novel approach to catch intra-day variability in RQ AGN simultaneously in X-ray and mm-waves. Indeed, the dramatic X-ray variability on short time scales known to exist in RQ objects should definitely help to look for correlated variability and to properly test the physical connection between the two bands. We present in this paper the result of the simultaneous NOEMA/NuSTAR campaign on the RQ AGN MCG+08-11-11, the main result of this paper being the detection, for the very first time, {of fast intraday variability in, and simultaneous increase between,} the mm and X-ray bands.
%The paper is written as follows. In Sect. \ref{targetselec} we present our target selection. In Sect. \ref{datareduc} we detail the data treatment and analysis of the NOEMA and NuSTAR observations. We then present and discuss our results in Sect. \ref{results} before concluding in Sect. \ref{discussion}. 

%%%%%%%%%%%%%%%%%%%%%%%%%%%%%%%%%%%%%%%%%%%%%%%%%%%%%%%%%%%%
\section{Target selection and observation strategy}
%%%%%%%%%%%%%%%%%%%%%%%%%%%%%%%%%%%%%%%%%%%%%%%%%%%%%%%%%%%%
\label{targetselec}
%The three sources of our sample are NGC 7469, NGC 5506 and MCG+08-11-11. These (using the fluxes from \cite{behar15,behar18}),(e.g. \citep{mat06}) 
For this project we selected bright X-ray AGN that are also bright ($\sim$5-10 mJy) in the mm range, show short ($\sim$1h) time scale variability in X-rays and are visible by NOEMA. Our two first targets were NGC 7469 and NGC 5506 that were observed in X-rays with {\it XMM-Newton}. Their unsuccessful campaigns are presented in the Appendix, where we also show a similar failed attempt with the JVLA and Chandra for Ark\,564.\\ %while the last one was observed in X-rays with NuSTAR. We were unlucky for the first two objects because either the source was not variable during the X-ray observation or the weather conditions were too bad at the NOEMA site for mm observations. A short description of the corresponding XMM/NOEMA campaigns  \\

The observations of MCG+08-11-11 (z=0.02) benefited from this experience and are the subject of this letter. MCG+08-11-11 is a well known and bright X-ray RQ AGN with a supermassive black hole mass of  $M_{BH} = (2.0\pm0.5)\times10^{7} M_{\odot}$ (e.g. \citet{bentz15} and references therein). It is among the few RQ AGN that have been observed at mm wavelengths with the ATCA and CARMA telescope arrays \citep{behar15,behar18}. It is bright ($\sim$5-10 mJy) at 95\, GHz, it shows short ($\sim$1h) time scale variability in X-rays with variations of $\sim$20\% in a few hours (e.g. \citealt{mat06}) and it was visible by NuSTAR and NOEMA for more than 12 hours at night at the beginning of the NOEMA winter period (December). 

The log of the observations is reported in Table \ref{logobs}. The NOEMA/NuSTAR observations of MCG+08-11-11 lasted $\sim$13h and were performed during the winter night 18-19 of December 2021.
{The details of the data reduction of both instruments are presented in Sect. \ref{datareduc} of the Appendix.}

\begin{figure*}[t]
    \centering
    \includegraphics[width=0.6\textwidth]{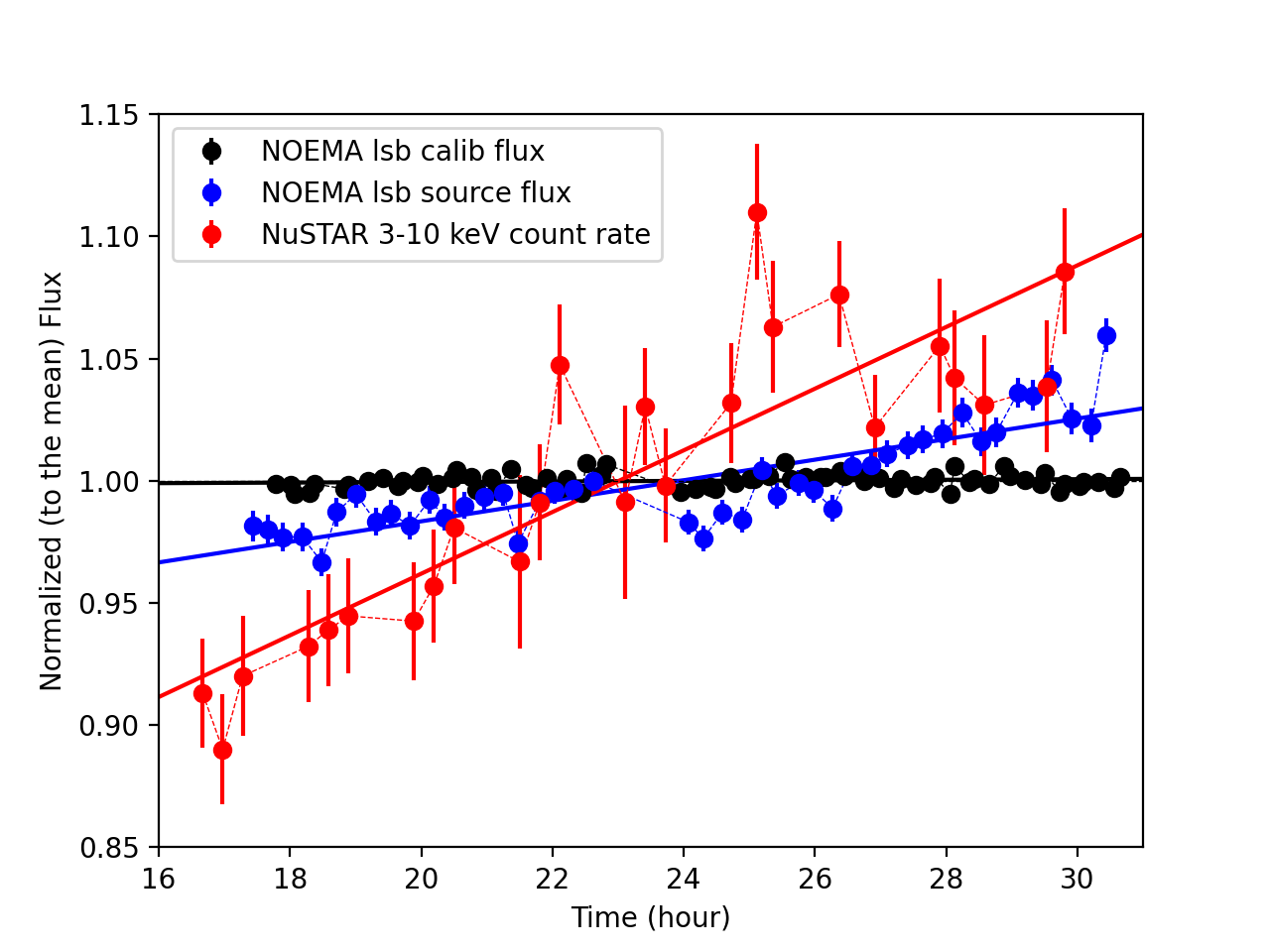}
%\includegraphics[width=0.45\textwidth]{fig1_210GHz.png}
%   \end{tabular}
     \caption{MCG+08-11-11 radio/X-ray variability. The black, blue and red circles/dashed lines are the NOEMA calibrator, the NOEMA LSB and 3-10 keV NuSTAR  light curves normalised by their mean. The solid lines are the corresponding linear best fits (see Sect. \ref{results}).  \label{fig2}}
 \end{figure*}

%%%%%%%%%%%%%%%%%%%%%%%%%%%%%%%%%%%%%%%%%%%%%%%%%%%%%%%%%%%%
\section{Results}
%%%%%%%%%%%%%%%%%%%%%%%%%%%%%%%%%%%%%%%%%%%%%%%%%%%%%%%%%%%%
\label{results}
The light curves, normalised to their mean, of the NOEMA LSB flux density at $\sim$100 GHz and the NuSTAR 3-10 keV X-ray count rate of MCG+08-11-11 are reported in Fig. \ref{fig2} in blue and red respectively. The NOEMA and NuSTAR time bins are of the order of 20 minutes. The NuSTAR and NOEMA normalised light curves clearly increase during the observations (even taking into account the small residual variability found in the calibrator, see Sect. \ref{calib}), while the one of the NOEMA calibrator remains constant. This is confirmed by linear fits of each normalised data sets with best fit values for the line slope equal to $(4.2\pm0.4)\times 10^{-3}$ hour$^{-1}$ and $(1.2\pm0.1)\times 10^{-2}$ hour$^{-1}$  for the NOEMA LSB and NuSTAR respectively. Both slopes are positive and significantly different from 0. This confirms the variability of the source in the two bands. This is the first time that mm variability on hour timescales is observed in a RQ AGN. 

\begin{figure}[t]
    \centering
    \includegraphics[width=\columnwidth]{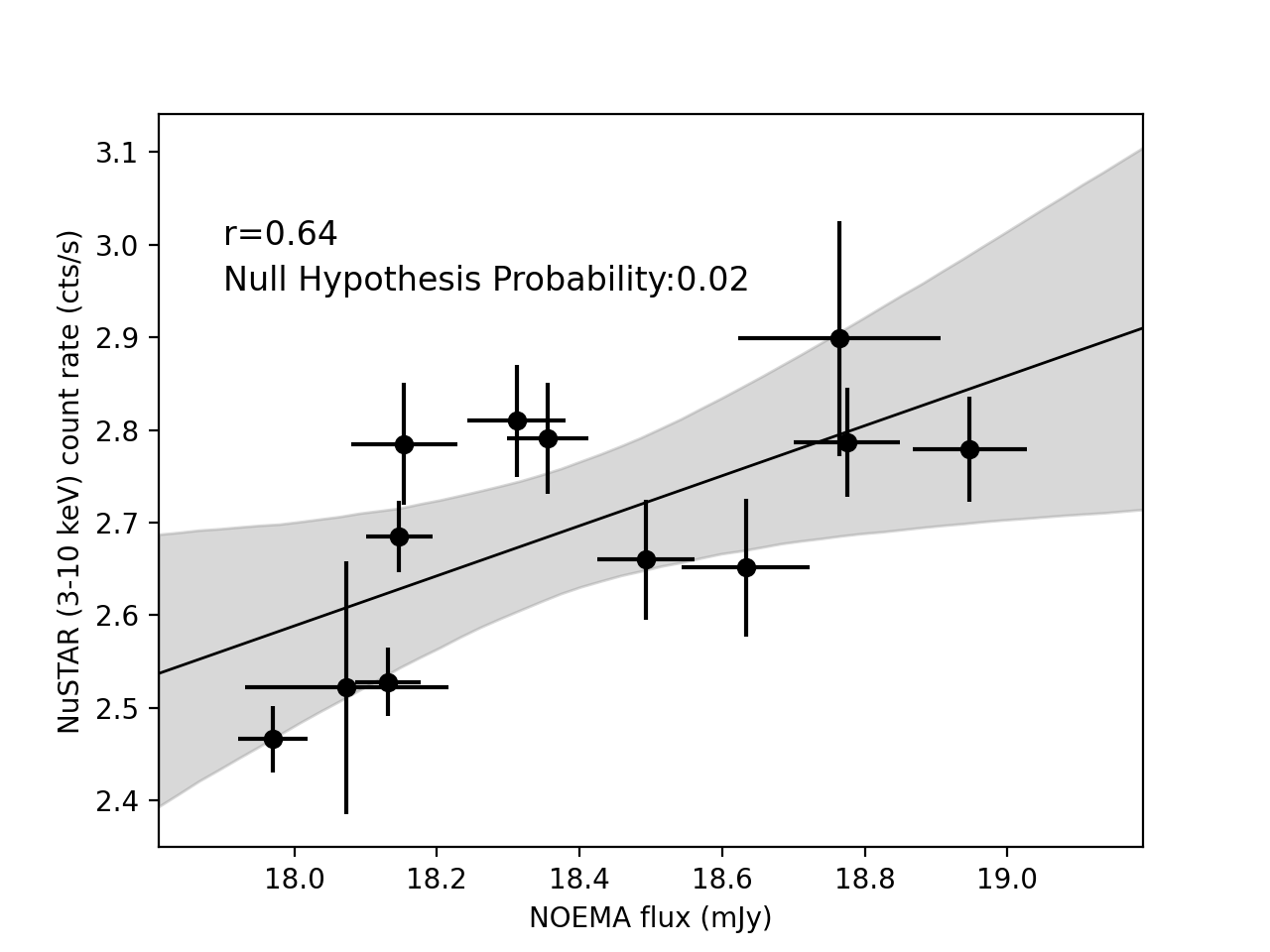}
%\includegraphics[width=0.45\textwidth]{fig1_210GHz.png}
%   \end{tabular}
     \caption[The weighted mean NuSTAR count rates vs NOEMA fluxes. The black solid line/grey shadowed area represent the best-fitting linear correlation and its 90\% confidence bands computed using the bayesian linear fitting {\sc python} package {\sc LINMIX}. The Spearman correlation coefficient $r$ and its corresponding p-value are indicated on the figure.]{{The weighted mean NuSTAR count rates vs NOEMA fluxes. The black solid line/grey shadowed area represent the best-fitting linear correlation and its 90\% confidence bands computed using the bayesian linear fitting {\sc python} package {\sc LINMIX}\footnote{https://github.com/jmeyers314/linmix}. The Spearman correlation coefficient $r$ and its corresponding p-value are indicated on the figure.}}\label{correl}
\end{figure}

Moreover, the fact that we detect the same variability behaviour (both mm and X-ray emission are increasing during the observation) is remarkable on such short timescales. {We apply the classical Spearman's rank correlation test on the mm and X-ray light curves. To do so, we need to take into account the gaps in the NuSTAR and NOEMA light curves and defined time zones where both instruments observed simultaneously (see Fig. \ref{GTIlightcurve}). Then we computed the weighted mean of the mm and X-ray light curves in these time zones. These weighted mean fluxes are reported in Fig. \ref{correl}. We find a Spearman correlation coefficient $r=0.64$ and its corresponding p-value of 0.02 indicating a quite strong correlation. The trueness of the causality between the two wavebands is admittedly questionable however given the short time interval on which the increasing trend is observed. We also searched for lags between the different light curves using the {\sc PyCCF}\footnote{https://bitbucket.org/cgrier/python\_ccf\_code/src/master/} {\sc python} code \citep{peterson98,sun18}. However this search was inconclusive with a flat cross-correlation function and a lag consistent with 0}.\\

%%%%%%%%%%%%%%%%%%%%%%%%%%%%%%%%%%%%%%%%%%%%%%%%%%%%%%%%%%%%
%\subsection{MCG+08-11-11}
%%%%%%%%%%%%%%%%%%%%%%%%%%%%%%%%%%%%%%%%%%%%%%%%%%%%%%%%%%%%
\begin{table*}
\begin{center}
\begin{tabular}{ccccccccc}
\hline
 $\left < F_X \right >$ & $\sigma_{F_X}$ & $F^X_{var}\pm \sigma_{F^X_{var}}$ & $\left < F_{mm} \right >$ & $\sigma_{F_{mm}}$  & $F^{mm}_{var}\pm\sigma_{F^{mm}_{var}}$  & $\left < F_{calib}\right >$ & $\sigma_{F_{calib}}$ & $F^{calib}_{var}\pm\sigma_{F^{calib}_{var}}$  \\%& $\chi^2$/dof & p-value \\
 (cts s$^{-1}$) & (cts s$^{-1}$) & (per cent) & (mJy) & (mJy) & (per cent) & (mJy) & (mJy) & (per cent) \\
 \hline
 %NGC 5506 & Total & 10.55$\pm$0.02 & 0.96 & 8.2$\pm$0.2 & 16.76$\pm$0.05 & 2.81 & 27.9$\pm$0.2 & 243.7$\pm$0.4 & 68.0& 16.5$\pm$0.3\\ %5793/82& 0.0\\
% NGC 7469 & Total & 22.29$\pm$0.03 & 2.15& 9.3$\pm$0.1& 8.79$\pm$0.03& 0.49& 5.2$\pm$0.3&544.6$\pm$0.3& 29.7&5.4$\pm$0.1 \\ %267/41&0.0\\
 % & $<$ 8h & 24.08$\pm$0.05 & 0.61& 0$^a$& 8.71$\pm$0.03& 0.24& 2.3$\pm$0.3&552.3$\pm$0.1&6.5&1.2$\pm$0.1 \\ %83/24&$10^{-8}$\\
 % & $>$ 8h & 21.44$\pm$0.03 & 2.09& 9.4$\pm$0.2& 8.90$\pm$0.05& 0.69& 7.4$\pm$0.6 & 533.1$\pm$0.8 & 43.7 &8.2$\pm$0.1\\%159/16&0.0\\
 2.77$\pm$0.01 & 0.77 & 7.0$\pm$1.5& 18.33$\pm$0.02  & 0.4 & 2.0$\pm$0.1 & 692.3$\pm$0.1 & 2.1 &0.29$\pm$0.01\\ %657/43&0.0\\
 \hline
 \end{tabular}
 \end{center}
 \caption{Variability properties of X-ray (3-10 keV), mm (100 GHz) and NOEMA calibrator light curves of the MCG+08-11-11 observations: weighted
mean flux $\left < F_i \right >$ (i=$X$, $mm$ or $calib$), standard deviations $\sigma_{F_i}$, and fractional variability amplitudes  $F^i_{var}$ including systematics error of 0.3\% for the NOEMA light curve. 
 %$^a$ When $\left < \sigma_i\right >$ exceeds $\sigma_{F_i}$ , $F^i_{var}$ is not well defined, and a value of zero is given. 
 \label{fvartab}}
 \end{table*}
%%%%%%%%%%%%%%%%%%%%%%%%%%%%%%%%%%%%%%%%%%%%%%%%%%%%%%%%%%%%
\subsection{Variability estimates}
%%%%%%%%%%%%%%%%%%%%%%%%%%%%%%%%%%%%%%%%%%%%%%%%%%%%%%%%%%%%
To better quantify the variability of the light curves, and following \cite{behar20},  we have reported in Tab. \ref{fvartab} the weighted mean 2-10 keV count rates (combining the two NuSTAR detectors), $\left < F_X \right >$, the  weighted mean mm flux densities, $\left < F_{mm} \right >$, with their errors (including 0.3\% of systematics, see Sect. \ref{calib}) and their respective standard deviation $\sigma_{F_X}$ and $\sigma_{F_{mm}}$. %The  $\chi^2$ obtained by fitting the X-ray and mm light curves with respect to a constant are also listed with the associated p-values (the lower p the larger the possibility of a true variability). 
We have also computed the fractional variability amplitudes:
\begin{equation}
F^X_{var} = \sqrt{\frac{\sigma_{F_X}^2-\left < \sigma_X^2\right >}{\left < F_{X}\right >^2}} \mbox{ and }
F^{mm}_{var} = \sqrt{\frac{\sigma_{F_{mm}}^2-\left < \sigma_{mm}^2\right >}{\left < F_{mm}\right >^2}}
\end{equation}
%\begin{eqnarray}
%F^X_{var} &=& \sqrt{\frac{\sigma_{F_X}^2-\left < \sigma_X^2\right >}{\left < F_{X}\right >^2}}\\
%F^{mm}_{var} &=& \sqrt{\frac{\sigma_{F_{mm}}^2-\left < \sigma_{mm}^2\right >}{\left < F_{mm}\right >^2}}
%\end{eqnarray}
where $\left < \sigma_X^2\right >$ and $\left < \sigma_{mm}^2\right >$  are the mean of the squared of the flux errors. The error on the fractional variabilities, $\sigma_{F^X_{var}}$ and $\sigma_{F^{mm}_{var}}$ , are given in Eq. 2 of \cite{behar20} and they have been also reported in Tab. \ref{fvartab}. For comparison, we have also reported in this table the weighted mean flux density $\left < F^{calib}\right >$ with its error, the respective standard deviation $\sigma_{F_{calib}}$ and the fractional variability amplitude $F^{calib}_{var}$ with its error for the NOEMA calibrator.

The value of $F^{calib}_{var}=0.29\pm0.01\%$ shows that the NOEMA photometric accuracy is much better than 1\% during this observation. In comparison the  fractional variability amplitude of the source at 100 GHz, $F^{mm}_{var}=2.0\pm0.1\%$, confirms intraday variability of a RQ AGN at mm wavelength. A few NOEMA data points even suggest variability on hour timescale. This corresponds, in term of light travel time, to a size of tens of $R_g$ for a $2\times 10^7$ $M_{\odot}$ black hole. This indicates that at least a few percent of the mm emission of the source is coming from such small regions. On the other hand, while the increasing trend all along the monitoring looks stronger in X-rays than in the mm, the 3-10 keV fractional variability amplitude of the source is of about 7\%, due to the larger X-ray error bars. \\

%We can interprete these results in two ways. The first one is that the mm emission is a combination of a dominating (in term of flux) but constant component, e.g. signature of a large scale emitting regions like a star forming region or a weak/small scale jets , and a variable one that produces the 3-4\% increase ($\sim$0.5mJy) observed during our $\sim$10h monitoring. Assuming these $\sim$0.5mJy correspond to the significant changes of the variable source flux,  we can estimate an upper limit on the emitting region size, i.e. $R< R_{max} = cT_{var}$ with $T_{var}\simeq10$hr. This gives $R_{max}\simeq10^{15}$ cm ($\sim$400 $R_g$ for the BH mass of MCG+08-11-1). On the other hand from Eq. \ref{eq1}, the size of the emitting region $R_{sync}=6\times 10^{14} B_G^{1/4}$ (we use the redshift of MCG+08-11-11 of 0.02).
%This gives thus an upper limit on the magnetic field strength $B_G< 10$ G.

%The second possibility is that the 3-4\% variability in 10hr corresponds to the variability of a unique region that radiates on average about 18 mJy. Its doubling times scale would correspond then to $\sim$250 hr. The same reasonning as above, but with these numbers, then constrain the maximal size of the emitting region to be  $R_{max}\simeq 3\times10^{16}$ cm ($\sim$ 10 thousands $R_g$) and a magnetic field strength $B_G< 10^4$ G.\\

\begin{figure}[t]
    \centering
    \includegraphics[width=\columnwidth]{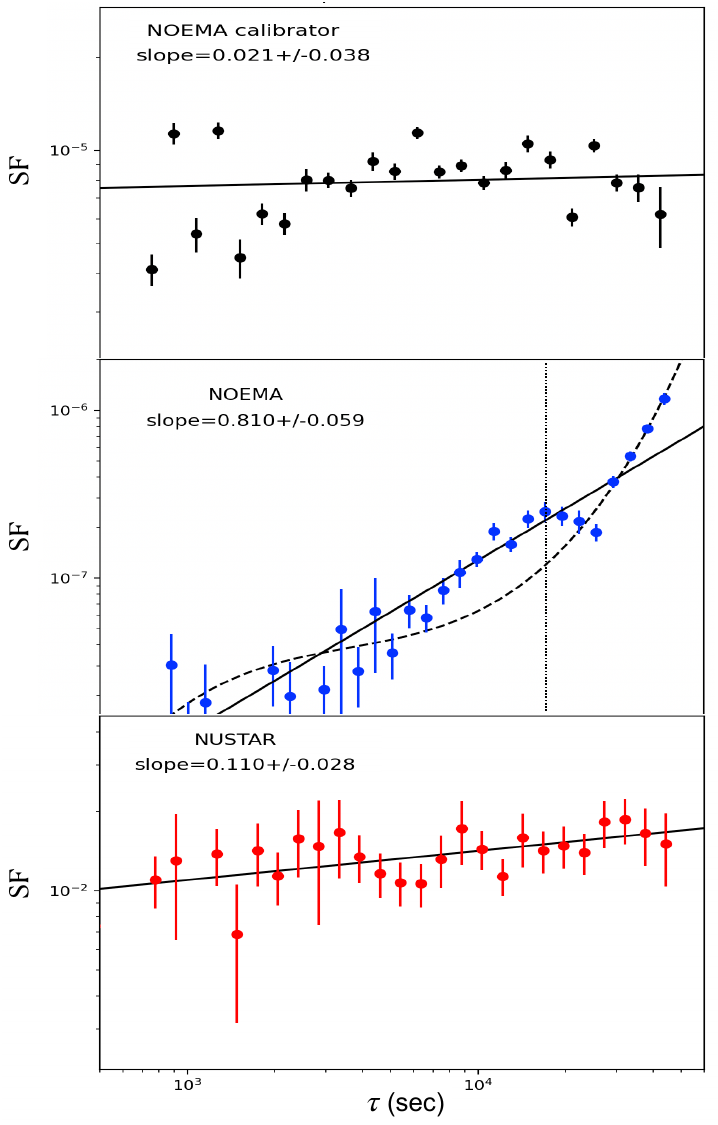}
%\includegraphics[width=0.45\textwidth]{fig1_210GHz.png}
%   \end{tabular}
     \caption{Structure function for the NOEMA calibrator (top), the NOEMA (middle) and NuSTAR (bottom) observations of MCG+08-11-11. The solid line is the best linear log-log fit, the corresponding slopes (with the corresponding errors) are indicated in each figure. The vertical dotted line in the middle plot indicates a local maximum around $\tau=2\times 10^4$ s. {The dashed line corresponds to the cubic polynomial best fit (in log scale) which is used to plot the data/model ratio in Fig. \ref{excessPSD}.\label{figStrucFun}}}
 \end{figure}

%%%%%%%%%%%%%%%%%%%%%%%%%%%%%%%%%%%%%%%%%%%%%%%%%%%%%%%%%%%%
\subsection{Structure function}
%%%%%%%%%%%%%%%%%%%%%%%%%%%%%%%%%%%%%%%%%%%%%%%%%%%%%%%%%%%%
\label{SFsection}
An interesting method to better quantify time variability when we have a quite small number of measurements (limiting the use of standard Fourier analysis) is via a structure function (SF) analysis (e.g. \citealt{simonetti85,hughes92,gliozzi01}). The shape and extrema of the structure function can indeed reveal the range of time scales that contribute to the variations in the data set (see, e.g., \citealt{paltani99}). 
There are different definitions of the structure function in the literature (e.g., \citealt{simonetti85,hughes92,diclemente96,vagnetti11,middei17}). We have chosen the following one (we have checked that the results do not depend qualitatively on the structure function expression):
\begin{equation}
 \mbox{SF}(\tau) = \left <\left [F(t + \tau)-F(t)\right ]^2\right >-\sigma_{noise}^2
 \end{equation} 
 where $F(t)$ and $F(t + \tau)$ are two measures of the flux, $\tau$ is the time lag between these two flux measurements and the $\left <\right >$ means that an average is computed within an appropriate bin of time lag around $\tau$. The term  $\displaystyle \sigma^2_{noise} = \left <\sigma^2_{F(t)} + \sigma^2_{F(t + \tau)}\right >$ is the quadratic contribution of the photometric noise to the observed variations. The computation of the errors of the structure function are detailed in Sect. \ref{errSF}. \\

The structure functions of the NuSTAR and NOEMA light curves of MCG+08-11-11 are reported in Fig. \ref{figStrucFun}. The structure function of the light curve of the NOEMA calibrator is also reported at the top of this figure. The linear (in log space) best fit of the structure functions are over-plotted in black solid line and the corresponding slope is indicated on each plot. In the  case of the NOEMA calibrator, the slope of the structure function is consistent with 0, with an average variability fraction $\sqrt{\mbox{SF}^{calib}_{mm}}/\left < F_{mm}\right >$ of less than 0.5\% on the whole range of time lags explored. The slope is however significantly different from 0 for the NuSTAR and NOEMA light curves. In the case of NuSTAR, the significance is just above 3$\sigma$ (due to the large error bars) while it is $>10\sigma$ for NOEMA. 

More interestingly, we observe a local maximum of the NOEMA structure function around $\tau$$\sim$2$\times 10^4$ s (indicated by the vertical doted line in Fig. \ref{figStrucFun}) suggesting a typical variability timescale of a similar order. {To make this local maximum more apparent, we have reported in red points in Fig. \ref{excessPSD} the ratio between the NOEMA SF and its cubic polynomial best fit (reported in black dashed line in the middle plot of Fig. \ref{figStrucFun}) obtained by ignoring the SF point in the time lag range [$7\times10^3-2.5\times10^4$] seconds.}

\begin{figure}
    \centering
    \includegraphics[width=\columnwidth]{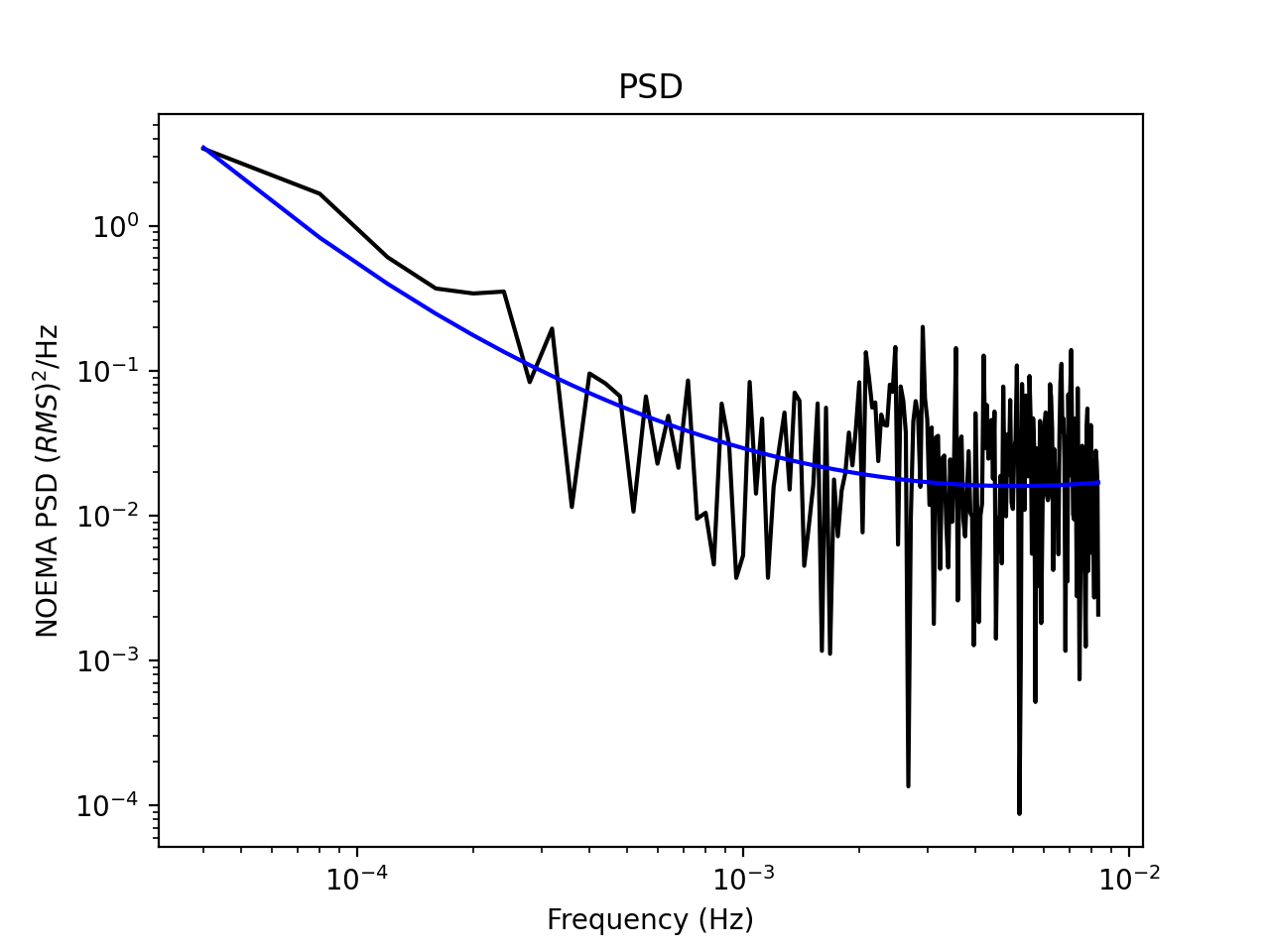}
%\includegraphics[width=0.45\textwidth]{fig1_210GHz.png}
%   \end{tabular}
     \caption{{PSD of the NOEMA light curve (black line). The quadratic (in log space) best fit is reported in blue.} \label{psdNOEMAnew}}
\end{figure}

\begin{figure}
    \centering
    \includegraphics[width=\columnwidth]{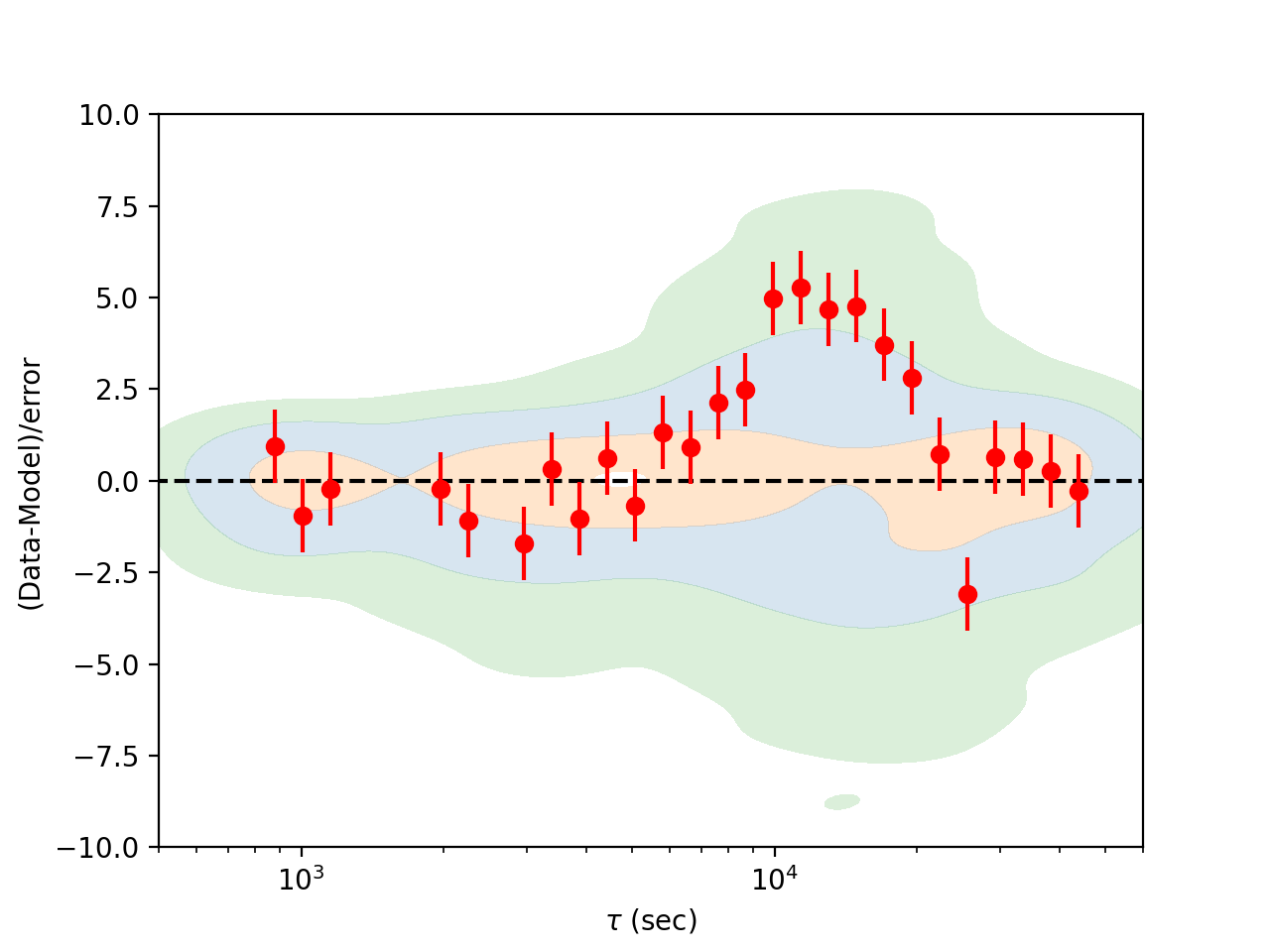}
%\includegraphics[width=0.45\textwidth]{fig1_210GHz.png}
%   \end{tabular}
     \caption{{The red points correspond to the residuals in term of sigma between the NOEMA SF data points and its cubic polynomial best fit reported in the middle plot of Fig. \ref{figStrucFun}. The colored area correspond to the 50\% (orange), 90\% (blue) and 99\% (green) percentile of the distribution of the same ratios but obtained with simulated data.}}\label{excessPSD}
\end{figure}

{We tried different tests to check the pertinence of this timescale in the NOEMA data. We first performed a Fourier analysis anyway. We computed the Power Spectral Density (PSD) of the NOEMA light curve using the astrophysical spectral-timing Python software package {\sc Stingray} v1.1.2\footnote{https://docs.stingray.science/index.html} \citep{huppenkothen2019a,huppenkothen2019b}. We used the {\sc AveragedPowerspectrum} class adapted to not-regularly-sampled light curves. It is reported in Fig. \ref{psdNOEMAnew}. The PSD statistics is quite poor at low frequency (which corresponds to the SF peak). This is expected since it is very close to the lowest possible frequency of the light curve. We do not detect any excess that would be the counterpart of the local maximum detected in the SF. 

Estimating the significance of such feature in a SF is not straightforward. We thus tried to reproduce such a local maximum via simulations. For that purpose we generated $10^4$ NOEMA light curves using the quadratic (in log scale) best fit of the PSD of the observed data (reported in blue in Fig. \ref{psdNOEMAnew}). This quadratic best fit was used as the PSD input to the light curve simulator {\sc python} package {\sc pyLCSIM}\footnote{https://pabell.github.io/pylcsim/html/code.html} \citep{pylcsim}. The phases being generated randomly in the process, most of the simulated light curves do not behaves like the observed one. Thus we selected the simulated light curves which are at less than 5 $\sigma$ from the observed one. For each of these selected simulated light curves we produced the corresponding SF following the procedure applied to the real data. An example of a simulated light curve and the associated SF are plotted in Fig. \ref{simulcsf} with the observed ones for comparison. Then we computed the ratio between the simulated SF and its best fit cubic polynomial. The orange, blue and green areas reported in Fig. \ref{excessPSD}  correspond to the  50\%, 90\% and 99\% percentile of the distribution of these simulated ratios. The observed local maximum peaks above the 90\% contour meaning that a similar maximum is reproduced in less than 10\% of the simulated SF. While this is by no mean an estimate of the significance of the observed peak (longer observations would be necessary for that), our procedure shows that this is not a generic feature which can be easily reproduced. }\\

%It is clear however that a linear best fit does not well reproduce the observed NOEMA PSD at low frequencies. A quadratic polynomial gives a much better fit in the entire frequency range (see the red line in Fig. \ref{psdNOEMA}). Applying the same methodology than above but using the quadratic polynomial best fit as PSD input to simulate the light curves, we now obtain the contours reported at the bottom of Fig. \ref{excessPSD}. The observed local maximum is now well inside the 95\% contour. This test suggests that there is a fraction of the the variability power present at low frequency (below 2-3$\times 10^{-4}$ Hz) which is the Fourier counterpart of the local maximum observed in the SF at 2$\times 10^4$ seconds.

{Assuming then this peak is real,} for the BH mass of MCG+08-11-11, its timescale corresponds to a light crossing time of 100-300 $R_g$. 
The value of the mm SF at this peak corresponds to an average variability fraction $\sqrt{\mbox{SF}_{mm}(\tau=2\times10^4)}/\left < F_{mm}\right >$ of about 2.5\%. 
Assuming the mm emission is optically thick, the variation of the observed flux is then related to the variation of the surface of the emitting region, i.e., to the squared of its typical size. An upper limit of this typical size can then be estimated assuming the emitting surface is homogeneous. Then 100\% variation of its flux would correspond to total emitting surface which is a factor $\sim$6-7 larger in radius. This puts the upper limit of the mm emitting region typical size of the order of $\sim$1300 $R_g$.

This is about the accretion disk to BLR scales, i.e., significantly larger than the size of the hot corona as deduced by, e.g., micro-lensing (e.g., \citealt{morgan2010,chartas2016}). Despite this difference in size, our results suggest that the two emitting regions are connected one with each other, indicating that the mm radio emitting region could be an extension of the hot corona, like an outflow/weak jet (see the discussion in \citealt{pane19}). Such outflow components are indeed expected as soon as a poloidal magnetic field component is present in the accretion flow (e.g., \citealt{beck08}). 

In this respect, using Eq. \ref{eq1b} the upper limit in size translates in an upper limit of the magnetic field of a few Gauss for the radio flux density of 18 mJy observed for MCG+08-11-11. This is an admittedly fairly loose constraint but this magnetic field strength is of the order to the one expected at a distance of 1300 $R_g$ for a magnetic field distribution which starts around a few $R_g$ in equipartition with radiation at Eddington luminosity around a supermassive black hole of $10^7 M_{\odot}$ (e.g., \citealt{ree84}).

It is also worth noting that the mm SF starts to re-increase after this local maximum, with a steeper slope, and with no evidence for flattening in agreement with longer (day) timescale variability observed in mm several RQ AGN (e.g., \citealt{doi11,baldi15,behar20}). In comparison, the NuSTAR structure function is rather flat down to the smallest time lags ($\tau<10^3$ sec), without apparent peculiar timescale, and with a variability fraction of about $\sqrt{\mbox{SF}_X}/\left < F_{X} \right > >$4\% on the whole range of time lags explored.

\section{Concluding remarks}
%%%%%%%%%%%%%%%%%%%%%%%%%%%%%%%%%%%%%%%%%%%%%%%%%%%%%
\label{discussion}
We report in this letter the results of a strictly simultaneous NOEMA/NuSTAR observation of $>$10 hr of the RQ AGN MCG+08-11-11. For the very first time we observe intraday (a few hours) variability of a few percent of the 100 GHz emission of a RQ AGN. Moreover, the mm and X-ray light curves show a similar increasing trend all along the observation.  This fast variability in the mm band and the apparent correlation with the X-rays clearly suggest a strong physical link between the two emitting regions.

A structure function analysis suggests a typical mm-wave variability of 2-3\% on a timescale around $2\times 10^4$ seconds, which translates in an upper limit of the size of the emitting regions $\sim$1300 $R_g$. These results indicate that the mm radio emitting region could be an extension of the hot corona, like an outflow/weak jet. 

New mm-X-ray observations are needed to support or rule-out different interpretations, especially if we are able to detect a delay in the mm and X-ray light curves which would put important constraints to study the mm-wave emission mechanism. Along these lines, an XMM/NOEMA campaign on NGC 4051 is expected during Winter 2023-2024.

%In addition to their physical impact to better understand the origin of the mm emission in RQ AGN, these new results also validate the selection criteria of the best RQ AGN candidates, as well as the best observation strategy: 
%1) the targets should be X-ray bright and strongly variable
%2) they should pass close to the zenith to allow for long visibility windows ($\sim$11h). % with NOEMA and 
%3) they must be simultaneously observed during several hours by X-ray instruments (e.g. XMM, Chandra or NuSTAR) during the winter period at the NOEMA site, the winter being the best time for the stability of the mm observations. 
%Moreover, instruments like XMM and Chandra would be preferred to NuSTAR to increase the statistics of the X-ray light curves. Along these lines, an XMM/NOEMA campaign on NGC 4051 is expected during Winter 2023-2024.

%%%%%%%%%%%%%%%%%%%%%%%%%%%%%%%%%%%%%%%%%%%%%%%%%%%%%
\begin{acknowledgements}
%%%%%%%%%%%%%%%%%%%%%%%%%%%%%%%%%%%%%%%%%%%%%%%%%%%%%
{The strict simultaneity, required between NOEMA and NuSTAR or XMM to be able to catch simultaneous variability in both bands, was perfectly conducted thanks to the efforts of the Science Operation Committee of the different observatories. The results presented in this letter would not have been possible without their precious help. Thanks a lot to all the SOC people of both instruments!}
Part of this work has been done thanks to the financial supports from CNES and the French PNHE.
The Technion group was supported in part by a Center of Excellence of the Israel Science Foundation (grant No. 1937/19). The scientific results reported in this article are based on observations with the XMM-Newton and  NuSTAR satellites and NOEMA ground based telescopes. This work is based on observations carried out under project numbers S21BI and W21BU with the IRAM NOEMA Interferometer. IRAM is supported by INSU/CNRS (France), MPG (Germany) and IGN (Spain).

%We would like to thank the referee for her/his careful reading of the manuscript and useful comments that help to improve its quality. This work is based on observations obtained with XMM-Newton, an ESA science mission with instruments and contributions directly funded by ESA Member States and the USA (NASA). We acknowledge financial support from the CNRS/INAF french/Italian PICS programme. POP acknowledges financial support from CNES and the French PNHE. POP and FU acknowledge support from the Italo/French Vinci programme. The research leading to these results has received funding from the European Union's Horizon 2020 Programme under AHEAD project (grant agreement n. 654215). SB and GM acknowledge financial support from the European Union Seventh Framework Programme (FP7/2007-2013) under grant agreement no. 31278. MD and MC acknowledges support from the ASI - INAF grant I/037/12/0
\end{acknowledgements}

%%%%%%%%%
\appendix
%%%%%%%%%
%%%%%%%%%%%%%%%%%%%%%%%%%%%%%%%%%%%%%%%%%%%%%%%%%%%%%%%%%%%%
\section{Observation and Data reduction}
%%%%%%%%%%%%%%%%%%%%%%%%%%%%%%%%%%%%%%%%%%%%%%%%%%%%%%%%%%%%
\label{datareduc}
%%%%%%%%%%%%%%%%%%%%%%%%%%%%%%%%%%%%%%%%%%%%%%%%%%%%%%%%%%%%
\subsection{NOEMA}
%%%%%%%%%%%%%%%%%%%%%%%%%%%%%%%%%%%%%%%%%%%%%%%%%%%%%%%%%%%%
\subsubsection{Observation}
%%%%%%%%%%%%%%%%%%%%%%%%%%%%%%%%%%%%%%%%%%%%%%%%%%%%%MGC+08-11-11 was observed with NOEMA with 9 antennas in the C configuration. The receivers were tuned with a LO frequency of 92 GHz, so that the lower sideband (lsb) covers 80.384-88.128 GHz, the upper sideband (usb) covers 95.872-103.616 GHz. We used only low spectral resolution spectral windows (2 MHz channel spacing).

The weather conditions for the NOEMA observation were excellent, with less than 1mm of precipitable water vapor. The phase stability was excellent with less than 15${^\circ}$ r.m.s. We used 3C454.3 for the bandpass calibration, and MWC349 for the flux calibration. Amplitude and phase calibrator was 0538+498, only 4 degrees away from MGC+08-11-11. Compared to standard NOEMA projects, we used a shortened calibration cycle of $\sim$13min to better track instrumental and weather variations.

The source was observed from 17.3h UT, 18 December 2021 to 06.5h UT, 19 December 2021 with a 1.2h gap when the source transits at high local elevation and is impossible to track. Data were reduced using the standard NOEMA pipeline. At the time of observation, a slight non-closure problem was affecting NOEMA, so baseline-based amplitude calibration was used. Any residual amplitude gain error due to the distance between calibrator and source is estimated to be $<1\%$.

The data were then exported in uv tables in small time chunks (separately for lower sideband (LSB) and upper sideband (USB)). Each uv table corresponding to one time step was then self-calibrated in phase, and a circular gaussian was fitted to the data in the (u,v) plane. The fitted fluxes and associated errors are then used to produce a light-curve.

%%%%%%%%%%%%%%%%%%%%%%%%%%%%%%%%%%%%%%%%%%%%%%%%%%%%%%%%%%%%
\subsubsection{Calibration}
%%%%%%%%%%%%%%%%%%%%%%%%%%%%%%%%%%%%%%%%%%%%%%%%%%%%%%%%%%%%
\label{calib}
The NOEMA calibrator light curve, with a time bin of $\sim$10 minutes and normalised to its mean flux, is reported in black in Fig.\,\ref{fig2}. A linear fit gives a best fit value for the slope of $(1.1\pm0.9)\times 10^{-4}$ hour$^{-1}$, i.e. almost consistent with 0 (at 1 $\sigma$), as expected for constant flux. The linear best fit is reported with a black solid line in Fig.\,\ref{fig2}. The result of a fit by a constant value (chi2/dof=1604/70) indicates however that some variability is not captured in the error bars. This is expected since the errors are computed from the thermal noise of the antenna and do not include their variations in gain. The standard deviation $\sigma_{F_{calib}}$ of the calibrator light curve allows us to estimate the systematic error related to this variation in gain. It is equal to 0.3\% (see Tab. \ref{fvartab}). It is then included in the total error of the NOEMA light curve (quadratically added to the thermal noise error) of the source.

%%%%%%%%%%%%%%%%%%%%%%%%%%%%%%%%%%%%%%%%%%%%%%%%%%%%%%%%%%%%
\subsection{NuSTAR}
%%%%%%%%%%%%%%%%%%%%%%%%%%%%%%%%%%%%%%%%%%%%%%%%%%%%%%%%%%%%
We calibrate and clean raw NuSTAR \citep{harrison13} data of MCG+08-11-11 using the NuSTAR Data Analysis Software (NuSTARDAS\footnote{see \href{http://heasarc.gsfc.nasa.gov/docs/nustar/analysis/nustar\_swguide.pdf}{http://heasarc.gsfc.nasa.gov/docs/nustar/analysis/nustar\_swguide.pdf}} %Perri et al., 2013) 
package v. 2.1.1). 
Level 2 cleaned data products were obtained with the standard {\it nupipeline} task while 3rd level science products (spectra and light curves) were computed with the {\it nuproducts} pipeline and using the calibration database 20211202. 
A circular region with a radius of 50\,arcsec centred on a blank area nearby the source was used to estimate the background. The extraction region for the source was selected using an iterative process that maximises the S/N similarly to what is described in \cite{pic04}. While we are mainly interested in the variability aspects in this paper, we had a quick look at the spectra. A simultaneous fit of the A and B module spectra in the 3-78 keV energy range with a cut-off power law + reflection component ({\sc{wabs*(cutoffpl+xillver)}} in {\sc{xspec}}) gives a reasonable fit with $\chi^2$/dof=3647/3743 corresponding to a Null hypothesis probability of $\sim$0.9. The best fit parameters are a power-law photon index $\Gamma=1.56\pm0.05$, a cutoff energy $E_c=20\pm2$\,keV and a reflection parameter $\simeq0.4$.  A more detailed spectral analysis will be done in a following paper. The present fit enables us to estimate the 2-10 keV luminosity (assuming $z=0.02$) of $L_{2-10\ \mbox{keV}}=4\times 10^{43}$ erg s$^{-1}$, and the corresponding bolometric luminosity (applying a X-ray bolometric correction factor $\kappa_X=10$ based on the empirical relation computed by \citealt{duras20}) of $L_{bol}=4\times 10^{44}$ erg s$^{-1}$, i.e., $\sim$15\% of the Eddington luminosity for a BH of $2\times10^{7}$ $M_{\odot}$.

%%%%%%%%%%%%%%%%%%%%%%%%%%%%%%%%%%%%%%%%%%%%%%%%%%%%%%%%%%%%
\section{Good time interval for correlation analysis}
\begin{figure}
    \centering
    \includegraphics[width=\columnwidth]{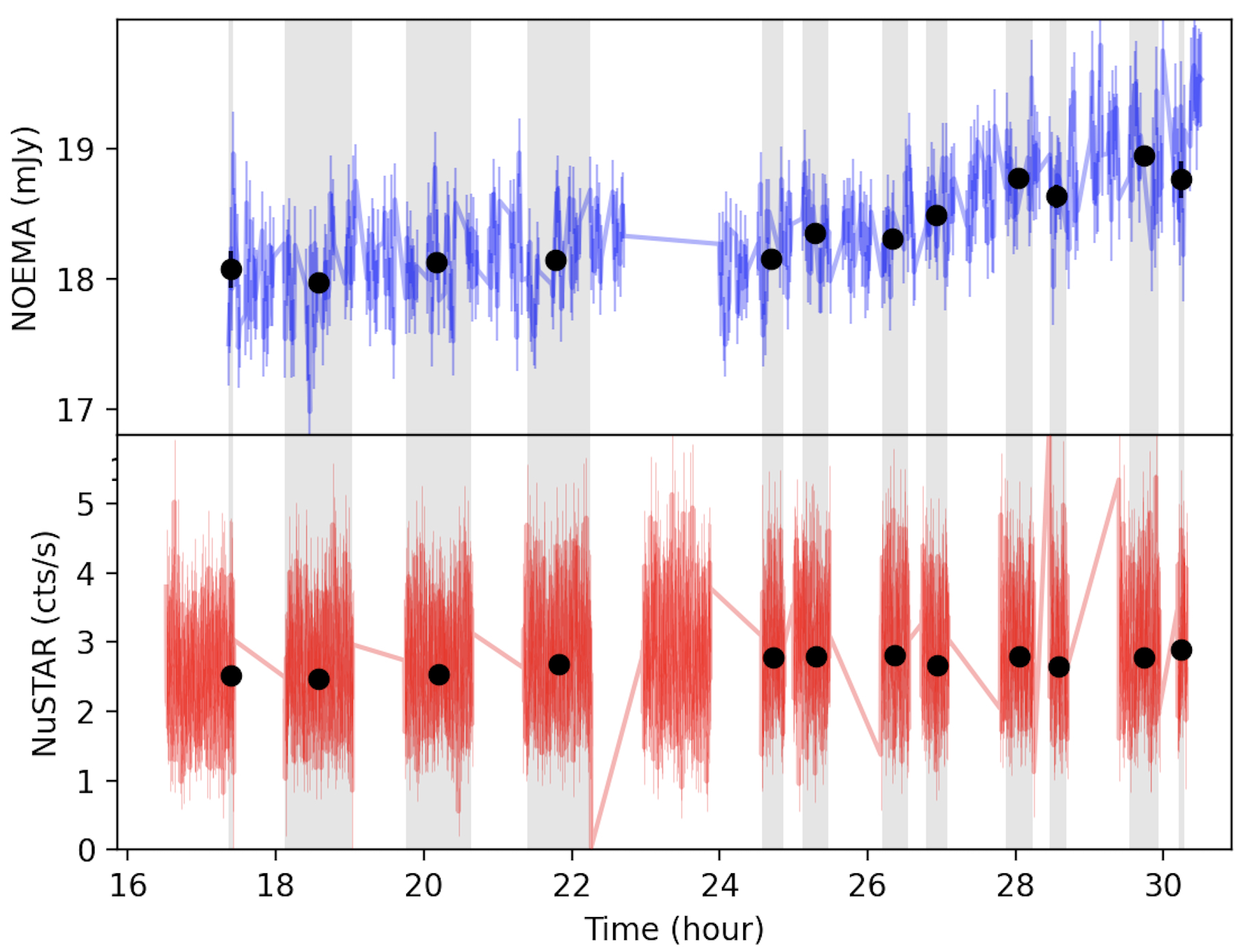}
%\includegraphics[width=0.45\textwidth]{fig1_210GHz.png}
%   \end{tabular}
     \caption{{Light curves of NOEMA (top, blue line) and NuSTAR (bottom, red line) with time bins of 60 and 6 sec. respectively. The 1.2h gap in the middle of the NOEMA light curve corresponds to the source transit at high local elevation where it becomes impossible to track. The gaps in the NuSTAR light curve correspond to Earth occultation and SAA passages. The grey areas correspond to the time ranges where both NOEMA and NuSTAR data are acquired simultaneously. The black points in each panel correspond to the weighted mean of each light curve in these different time ranges which are reported in Fig. \ref{correl}} \label{GTIlightcurve}}
\end{figure}

%%%%%%%%%%%%%%%%%%%%%%%%%%%%%%%%%%%%%%%%%%%%%%%%%%%%%%%%%%%%
{The light curves of NOEMA and NuSTAR, with a time binning of 60 and 6 sec respectively, are reported on top and bottom of Fig. \ref{GTIlightcurve}. The gaps in the NuSTAR light curve correspond to the time when the satellite was not acquiring valid data due to, e.g., Earth occultation or SAA passages. The gap in the middle of the NOEMA observation is when the source transited at high local elevation and was impossible to track.

The grey area on Fig. \ref{GTIlightcurve} correspond to the time zones where both NOEMA and NuSTAR data are acquired. The black points correspond to the weighted mean of each light curve in the different time zones. These weighted mean values have been used for the Spearman's rank correlation test discussed in Sect. \ref{results}.
}

%%%%%%%%%%%%%%%%%%%%%%%%%%%%%%%%%%%%%%%%%%%%%%%%%%%%%%%%%%%%
%\section{Power Spectral Density}
%%%%%%%%%%%%%%%%%%%%%%%%%%%%%%%%%%%%%%%%%%%%%%%%%%%%%%%%%%%%
%{\bf 
%We have computed the PSD of the NOEMA light curve using the astrophysical spectral-timing Python software package {\sc Stingray} v1.1.2 \footnote{https://docs.stingray.science/index.html} \citep{huppenkothen2019a,huppenkothen2019b}. We used the {\sc AveragedPowerspectrum} class adapted to not-regularly-sampled light curves like the NOEMA one. The resulting PSD is plotted in Fig. \ref{psdNOEMA}. It is normalized to the squared fractional RMS of the light curve. The linear and quadratic (in log space) best fits are reported in blue and red respectively.
%}
%\begin{figure}
%    \centering
%    \includegraphics[width=\columnwidth]{PSDNoema.jpg}
%%\includegraphics[width=0.45\textwidth]{fig1_210GHz.png}
%%   \end{tabular}
%     \caption{{\bf PSD of the NOEMA light curve (black line). The linear and quadratic (in log space) %best fits are reported in blue and red respectively.} \label{psdNOEMA}}
%\end{figure}

%%%%%%%%%%%%%%%%%%%%%%%%%%%%%%%%%%%%%%%%%%%%%%%%%%%%%%%%%%%%
\section{Simulated light curve and Structure function}
\begin{figure}
    \centering
    \includegraphics[width=\columnwidth]{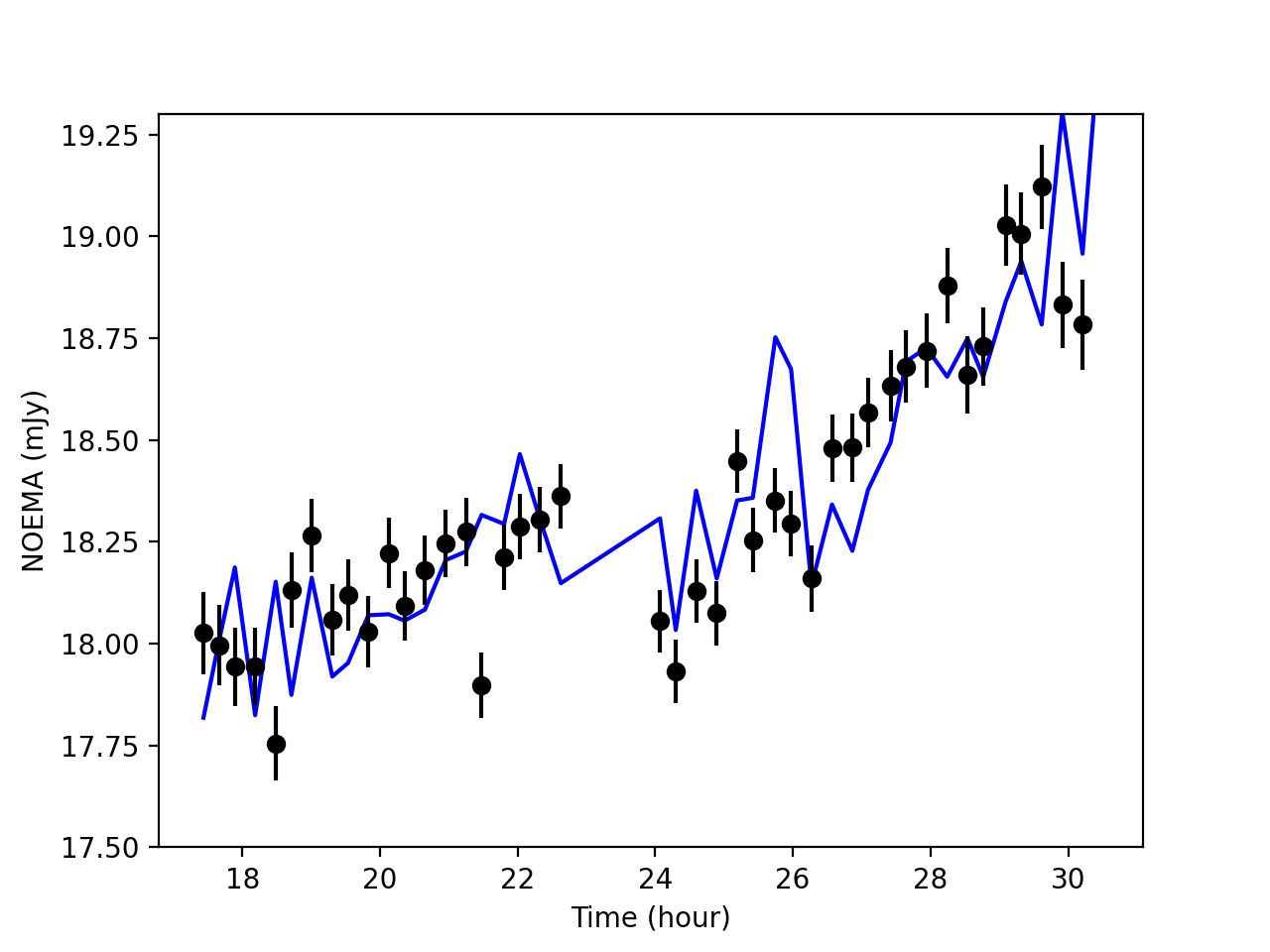}
    \includegraphics[width=\columnwidth]{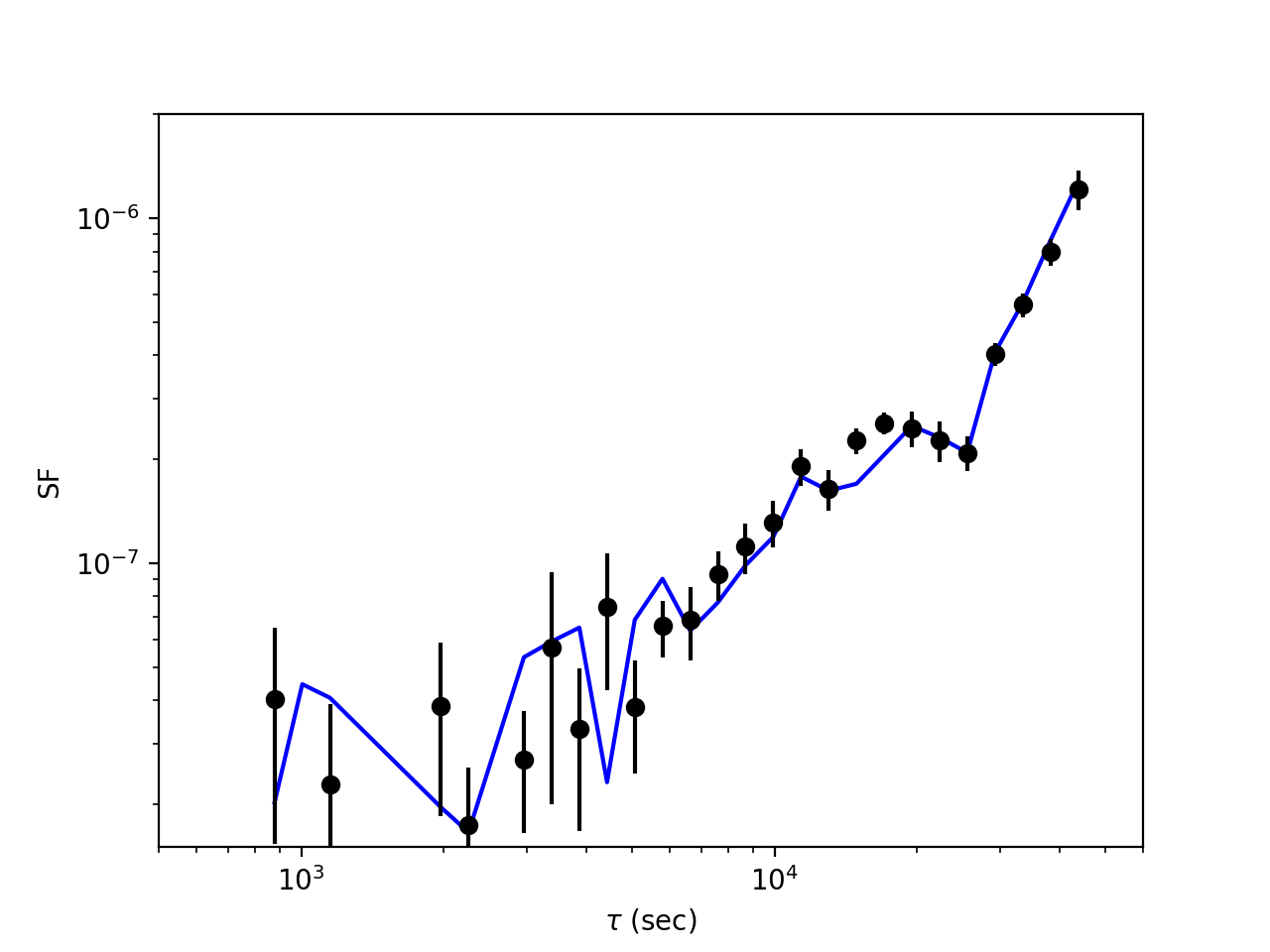}
%   \end{tabular}
     \caption{{Examples of simulated (blue lines) NOEMA light curve (top) and structure function (bottom).  The black points correspond to the observations.} \label{simulcsf}}
\end{figure}

%%%%%%%%%%%%%%%%%%%%%%%%%%%%%%%%%%%%%%%%%%%%%%%%%%%%%%%%%%%%
We have reported in Fig. \ref{simulcsf} an example of a simulated NOEMA light curve (top) and its corresponding Structure Function (bottom). The simulation procedure is explained in Sect. \ref{SFsection}.

%%%%%%%%%%%%%%%%%%%%%%%%%%%%%%%%%%%%%%%%%%%%%%%%%%%%%%%%%%%%
\section{The $R-B_G$ plane}
\begin{figure}
    \centering
    \includegraphics[width=\columnwidth]{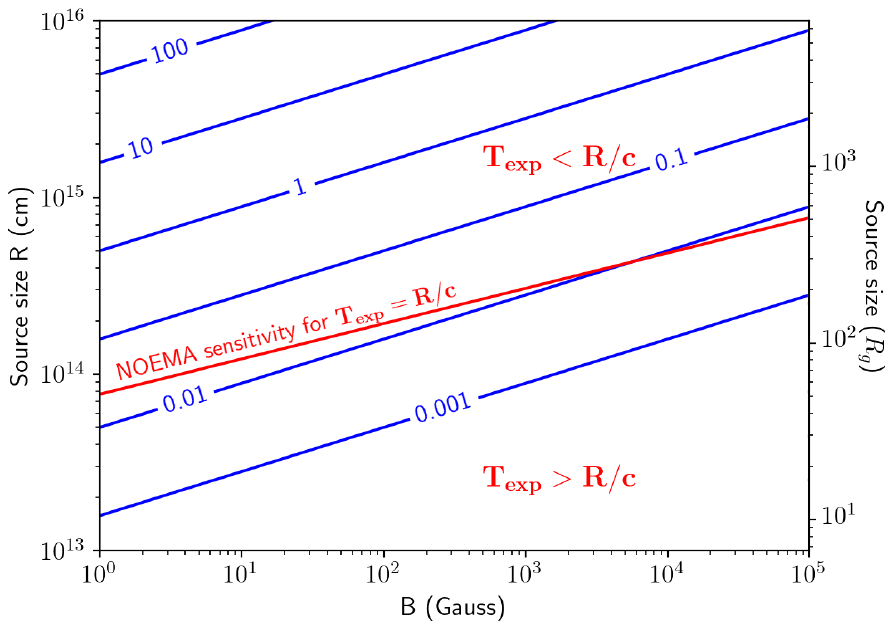}
%\includegraphics[width=0.45\textwidth]{fig1_210GHz.png}
%   \end{tabular}
     \caption{Contour plot (blue) of the radio flux density $F_{\nu}$ (in mJy) in the $R-B_G$ plane following the expression of $R$ given in Eq. \ref{eq1} for a frequency of 100 GHz and a source redshift z=0.02. The red line is the corresponding NOEMA sensitivity for an integration time equal to the source size light travel time $R/c$. Above this line, the required exposure time to detect a flux density  $F_{\nu}$ is lower than $R/c$. It is larger below.  The left y-scale is in cm unit while the right y-scale is in $R_g$ unit for a supermassive black hole mass of $10^7 M_{\odot}. $\label{fig1}}
\end{figure}

%%%%%%%%%%%%%%%%%%%%%%%%%%%%%%%%%%%%%%%%%%%%%%%%%%%%%%%%%%%%
The contour plot of the radio flux density $F_{\nu}$ in the $R-B_G$ plane is shown in Fig. \ref{fig1}. It results from Eq. \ref{eq1} and \ref{eq1b} assuming a redshift of 0.02 (which is the one of MCG+08-11-11) and for a frequency of 100\,GHz (which corresponds to the frequency of our NOEMA observation). The scaling in term of $R_g$ is indicated by the right y-scale of Fig. \ref{fig1}.

%%%%%%%%%%%%%%%%%%%%%%%%%%%%%%%%%%%%%%%%%%%%%%%%%%%%%%%%%%%%
\section{Structure function Errors}
%%%%%%%%%%%%%%%%%%%%%%%%%%%%%%%%%%%%%%%%%%%%%%%%%%%%%%%%%%%%
\label{errSF}
The errors on the Structure Function have been estimated by simulating $n_{trial}$ light curves from the observed ones. For that purpose, we have simulated light curves with the same number of data points than the observed ones, each data point being randomly distributed in a normal distribution centred on the observed flux and with a standard deviation equal to the observed photometric error. For each simulated light curve, we have computed the corresponding structure function SF$_i(\tau)$ ($i$ between 1 and $n_{trial}$). Then we deduced the structure function error at each $\tau$ from the standard deviation of the $n_{trial}$ measurements. A number $n_{trial}>$10 (we took 30) is generally sufficient to have a good error estimate of the structure function for the range of $\tau$ covered by our observed light curves.

\section{Previous tentatives}
The unsuccessful campaigns on NGC 5506, NGC 7469, and Ark\,564 are described below, were crucial for the development of this project. In addition to learning the best observing strategy for a successful campaign, we validated that with good weather conditions it was possible to reach a sensitivity of 12.5 $\mu$Jy/beam at 100 GHz, allowing us to catch a few \% variability for mJy sources. 
%%%%%%%%%%%%%%%%%%%%%%%%%%%%%%%%%%%%%%%%%%%%%%%%%%%%%%%%%%%%
\subsection{NGC 5506}
\begin{figure}[t]
   \includegraphics[width=\columnwidth]{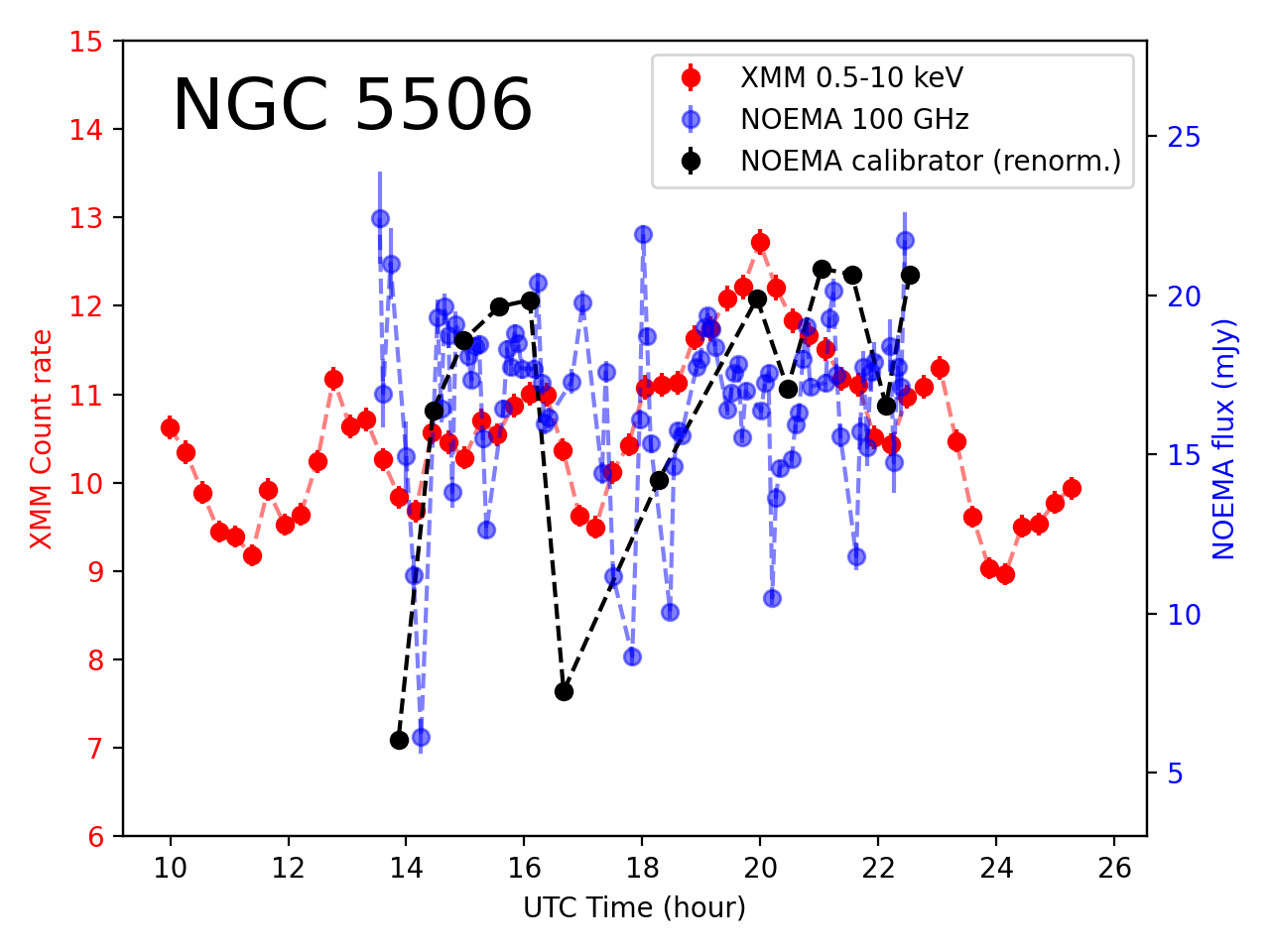}
 \caption{NOEMA (blue, right scale) and XMM X-ray (red, left scale) light curves NGC 5506 obtained during the first semester 2021. The NOEMA light curves (at 100 GHz) have a time binning of 4 min respectively. The XMM light curves have a time binning of 17 min. %The shaded gray area indicates the variation of the calibrators during each NOEMA pointings. 
The calibrator light curve is plotted in black (right scale). The calibration accuracy needed to constrain the mm variability cannot be reached during the entire observation for NGC 5506 due to bad weather conditions at the NOEMA site.\label{lc5506}}
%
%The systematics level deduced from the calibrator variation
%
\end{figure}

%%%%%%%%%%%%%%%%%%%%%%%%%%%%%%%%%%%%%%%%%%%%%%%%%%%%%%%%%%%%
NGC 5506 ($z$=0.006, $M_{BH}$ poorly known in between $10^6-10^8 M_{\odot}$, e.g. \citealt{matt2015})
was well detected by NOEMA at a few mJy. However, to fit with the XMM visibility window the observation had to be conducted during the summer. The weather conditions were very bad at the NOEMA site, preventing any good estimate of the mm variability, while clear X-ray variability occured during all the XMM pointing (see Fig. \ref{lc5506}). We have reported the light curve of the NOEMA calibrator in black in this figure. This observation demonstrates that to reach the conditions to detect weak  ($<$ 5-10 \%) variability in the mm on hour timescale, the summer seasons has to be avoided.

%%%%%%%%%%%%%%%%%%%%%%%%%%%%%%%%%%%%%%%%%%%%%%%%%%%%%%%%%%%%
\section{NGC 7469}
%%%%%%%%%%%%%%%%%%%%%%%%%%%%%%%%%%%%%%%%%%%%%%%%%%%%%%%%%%%%
\begin{figure}[t]
   \includegraphics[width=\columnwidth]{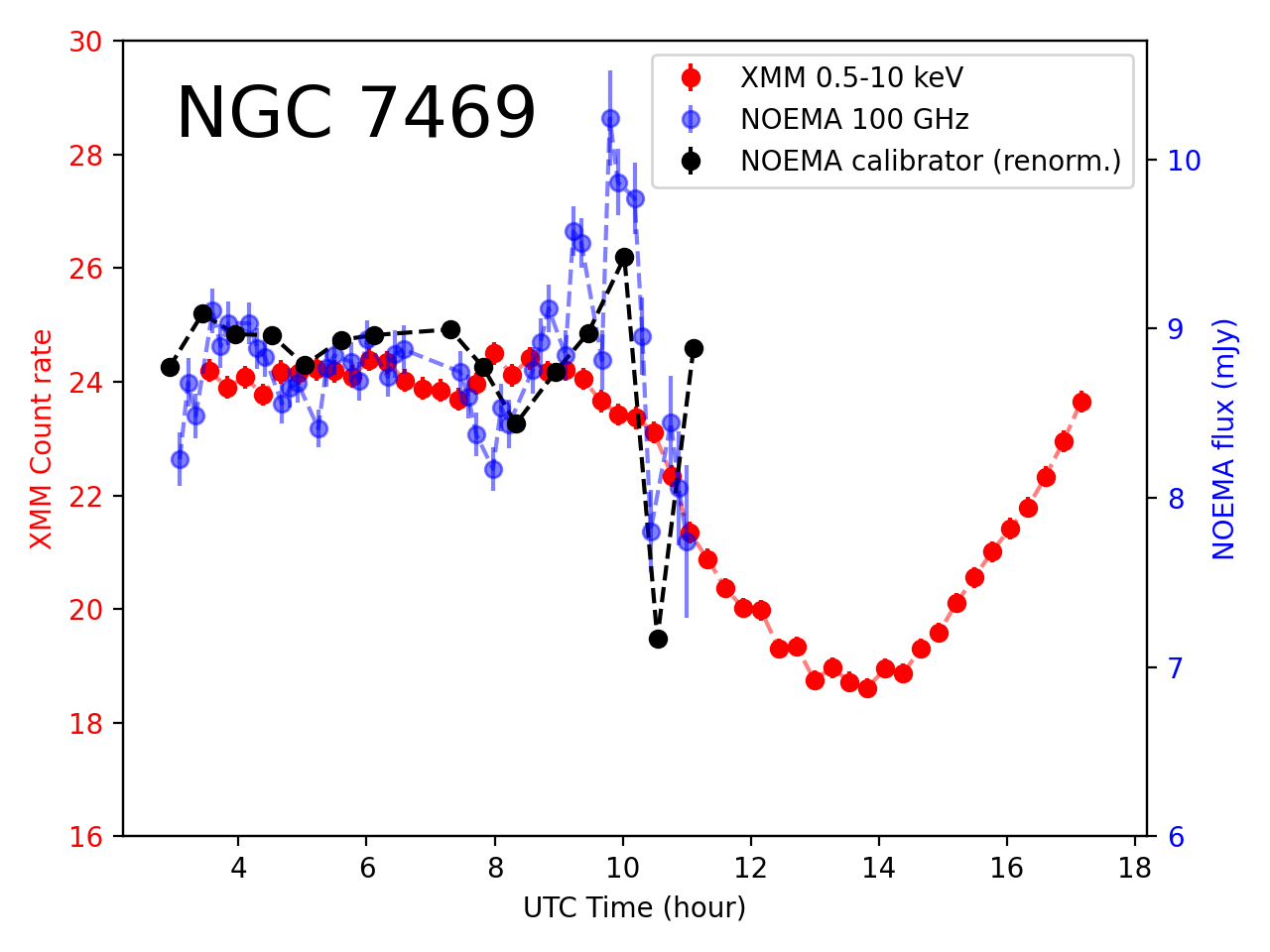}
\caption{NOEMA (blue, right scale) and XMM X-ray (red, left scale) light curves of NGC 7469 obtained during the first semester 2021. The NOEMA light curves (at 100 GHz) have a time binning of 7 min. The XMM light curves have a time binning of 17 min. %The shaded gray area indicates the variation of the calibrators during each NOEMA pointings. 
The calibrator light curve is plotted in black (right scale). The calibration accuracy needed to constrain the mm variability cannot be reached after 8-9h UTC.\label{lc7469}}
%%
%The systematics level deduced from the calibrator variation
%
\end{figure}

NGC 7469 ($z$=0.016, $M_{BH}=10^{6.9}M_{\odot}$, e.g. \citealt{peterson2014,bentz2015}) was also well detected by NOEMA at a few mJy. The weather was good at the NOEMA site. The NOEMA pointing lasted about 8 hours, as expected, while the XMM one, which cover entirely the NOEMA observation, was almost twice longer.  However the source stayed almost constant in X-ray during all the NOEMA coverage, its X-ray flux starting to significantly vary (after 8-9h UTC) when the lower source elevation and sunrise induce large amplitude gain variations at NOEMA (see Fig. \ref{lc7469}). These large amplitude gain variations produce variation in the light curve of the calibrator overplotted in black in this figure.

\section{Ark\,564}

\begin{figure}[t]
\begin{center}
   \includegraphics[width=\columnwidth]{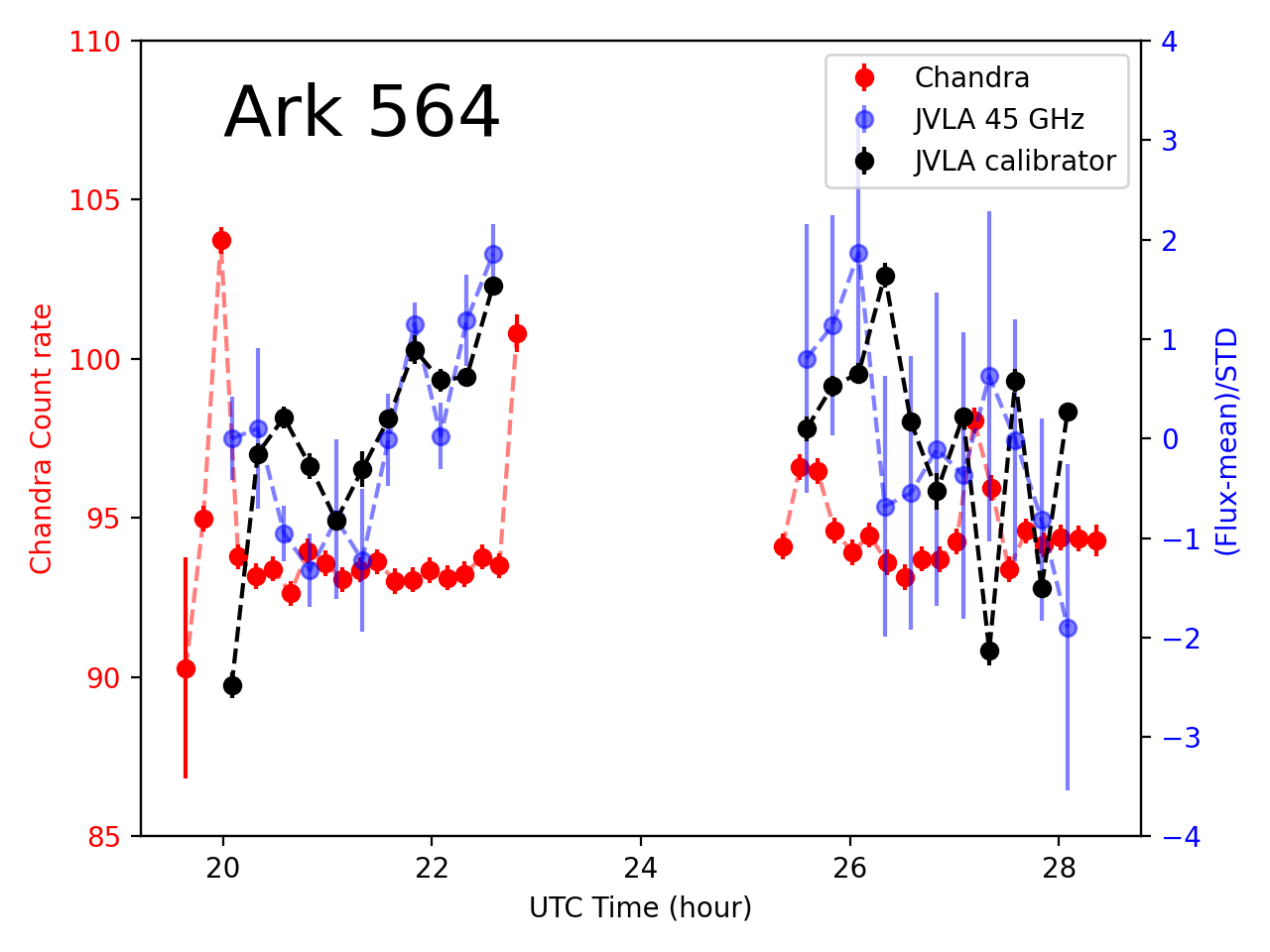}
\end{center}
\caption{JVLA 45\,GHz (blue, right scale) and Chandra X-ray (red, left scale) light curves of Ark\,564 over 3+3 hrs with a gap due to high sky elevation. Data are binned to 15\,min, about 12\,min on target for JVLA. The phase calibrator is plotted in black (right scale). No significant variability is detected at 45\,GHz, and only marginal X-ray variability can be seen.\label{lcArk564}. }
%%
%The systematics level deduced from the calibrator variation
%
\end{figure}

Ark\,564 ($z$=0.024, $M_{BH}=10^{6.2} M_{\odot}$, e.g., \citealt{zhang2006}) was chosen for a simultaneous X-ray radio campaign seeking intraday variability, due to its documented X-ray variability over hours, and its relatively low BH mass.
Chandra/LETG observed Ark\,564 in X-rays simultaneously with the JVLA at 45\,GHz, for 3+3 hours (gap again due to high sky elevation) on Dec. 20, 2019. 
At 45\,GHz, each observing scan is 4\,min on target and 1\,min on the phase calibrator. 
The light curves are shown in Fig\,\ref{lcArk564}. Unfortunately, no significant variability can be detected.

\end{document}